\documentclass[nofootinbib,aps,a4paper,letterpaper,superscriptaddress,
twocolumn,times]{revtex4}
\usepackage{amsmath}
\usepackage{amsfonts}
\usepackage{booktabs}
\usepackage{multirow}
\usepackage{siunitx} 
\usepackage{adjustbox}
\pdfoutput=1
\usepackage{graphicx}
\usepackage{color}
\usepackage{braket}
\usepackage{dcolumn}
\usepackage{bm,url}
  
\usepackage{amsfonts}
\usepackage[usenames,dvipsnames,svgnames]{xcolor}  
\usepackage{hyperref}   
\definecolor{oxfordblue}{rgb}{0.0, 0.13, 0.28}
\definecolor{burgundy}{rgb}{0.5, 0.0, 0.13}
\definecolor{darkolivegreen}{rgb}{0.33, 0.42, 0.18}
\definecolor{darkblue}{rgb}{0,0,0.5}
\definecolor{richcarmine}{rgb}{0.84, 0.0, 0.25}
\definecolor{darkblue}{rgb}{0,0,0.5}
\definecolor{bluer}{rgb}{0.00,0.50,0.75}{}
\hypersetup{colorlinks=true, citecolor=red, linkcolor=blue,
 urlcolor = magenta, filecolor=magenta}

\let\oldsqrt\sqrt
\def\sqrt{\mathpalette\DHLhksqrt}
\def\DHLhksqrt#1#2{%
\setbox0=\hbox{$#1\oldsqrt{#2\,}$}\dimen0=\ht0
\advance\dimen0-0.2\ht0
\setbox2=\hbox{\vrule height\ht0 depth -\dimen0}%
{\box0\lower0.4pt\box2}}

\newcommand{\sss}[1]{{\scriptscriptstyle{#1}}}

\newcommand{\uPl}{\mathrm{Pl}}

\newcommand{\usssPl}{\sss{\uPl}}

\newcommand{\Mp}{M_\usssPl}

\newcommand{\beq}{\begin{equation}}
\newcommand{\eeq}{\end{equation}}
\newcommand{\bea}{\begin{equation}\begin{aligned}}
\newcommand{\eea}{\end{aligned}\end{equation}}

\newlength{\wsingfig}
\newlength{\wdblefig}
\newlength{\wquadfig}
\newlength{\wtriplefig}
\newcommand{\Eq}[1]{Eq.~(\ref{#1})}

\newcommand{\Fig}[1]{Fig.~{\ref{#1}}}

\newcommand{\Sec}[1]{Sec.~\ref{#1}}

\newcommand{\bma}{\begin{pmatrix}}
\newcommand{\ema}{\end{pmatrix}}
\newcommand{\be}{\begin{equation}}
\newcommand{\beno}{\begin{equation*}}
\newcommand{\eeno}{\end{equation*}}

\def\doi{http://doi.org}

\usepackage{soul}

\usepackage{array}

\begin{document}

\title{Primordial black hole dark matter from ultra-slow-roll inflation in Horndeski gravity}

\author{Despina Totolou}
\email{dtotolo@auth.gr}
\affiliation{Department of Physics, Aristotle University of Thessaloniki, 54124 Thessaloniki, Greece}
\affiliation{Institute for Astronomy, Astrophysics, Space Applications and 
Remote Sensing, National Observatory of Athens, 15236 Penteli, Greece}

\author{Theodoros Papanikolaou}
\email{papaniko@upatras.gr}
\affiliation{Laboratory of Theoretical and Computational Physics, Department of Physics, University of Patras, 26504 Patras, Greece}
\affiliation{Scuola Superiore Meridionale, Largo San Marcellino 10, 80138 Napoli, Italy}
\affiliation{Istituto Nazionale di Fisica Nucleare (INFN), Sezione di Napoli, Via Cinthia 21, 80126 Napoli, Italy}
\affiliation{Institute for Astronomy, Astrophysics, Space Applications and 
Remote Sensing, National Observatory of Athens, 15236 Penteli, Greece}

\author{Emmanuel N. Saridakis}
\email{msaridak@phys.uoa.gr}
\affiliation{Institute for Astronomy, Astrophysics, Space Applications and 
Remote Sensing, National Observatory of Athens, 15236 Penteli, Greece}
\affiliation{Departamento de Matem\'{a}ticas, Universidad Cat\'{o}lica del 
Norte, 
Avda.
Angamos 0610, Casilla 1280 Antofagasta, Chile}
\affiliation{CAS Key Laboratory for Researches in Galaxies and Cosmology, 
Department of Astronomy, University of Science and Technology of China, Hefei, 
Anhui 230026, P.R. China}


\begin{abstract}
Primordial black holes (PBHs) provide a well-motivated non-particle candidate 
for dark matter, requiring an enhancement of curvature perturbations on small 
inflationary scales consistent with observational constraints. In this work  we 
study PBH production within Horndeski gravity, accounting for compatibility with the 
GW170817 constraint on the gravitational-wave (GW) speed and imposing a constant 
coupling to the Ricci scalar. Under these conditions, and assuming an inflaton field characterised by a canonical 
kinetic term and a smooth potential, the inflationary dynamics is 
controlled by the cubic Horndeski interaction.
By investigating standard functional forms of the latter we identify the specific kinetic structure that allows enhancement of the effective friction on the inflaton, thereby inducing a transient ultra-slow-roll phase embedded within a standard slow-roll evolution.
For representative parameter 
choices we find that pronounced amplifications in the scalar power spectrum are 
generated, leading to the formation of asteroid-mass PBHs with masses of order 
$\mathcal{O}(10^{-16})\,M_\odot$, which can account for a substantial fraction 
of the dark matter abundance, reaching $f_{\rm PBH}\simeq 0.9$, while satisfying 
current observational constraints. The resulting characteristic features in the scalar power 
spectrum also imply potentially observable scalar-induced gravitational-wave (SIGW)
signatures.
\end{abstract}

\maketitle
\section{Introduction}

According to recent cosmological observations by the Planck satellite, 
approximately $26\%$ of the total energy budget of the Universe consists of dark 
matter (DM), interacting predominantly through gravity with ordinary baryonic 
matter~\cite{Aghanim:2018eyx}. Despite the remarkable success of the standard 
cosmological model in describing a wide range of observations, the physical 
nature and origin of dark matter remain unresolved. The majority of proposed DM 
candidates are rooted in extensions of particle physics beyond the Standard 
Model~\cite{Bertone:2004pz, Feng:2010gw}, yet no conclusive experimental 
evidence for such particles has been established so far.

An alternative, non-particle candidate for dark matter is provided by primordial 
black holes (PBHs), which may have formed in the early Universe from the 
gravitational collapse of enhanced density fluctuations~\cite{Zeldovich:1967lct, 
Hawking:1971ei, Carr:1974nx}. Depending on the epoch of their formation, PBHs 
can span a very wide mass range, from the Planck scale, i.e. $\sim 
10^{5}\,\mathrm{g}$, up to supermassive values of $\mathcal{O}(10^{15}M_\odot)$. 
On the other hand, observational constraints on their abundance arise from a 
variety of probes, including the Cosmic Microwave Background (CMB), 
gravitational lensing, gravitational waves, accretion effects and large-scale 
structure (LSS) formation~\cite{Sasaki:2018dmp,Carr:2020gox, Green:2020jor,Escriva:2022duf}. Within 
certain mass windows, most notably at asteroid and sub-lunar masses, PBHs remain 
viable candidates that could account for a significant fraction, or even the 
totality, of the dark matter abundance.

PBH formation is commonly investigated in the context of inflation, where 
enhanced curvature perturbations on small scales may re-enter the horizon during 
radiation domination and collapse gravitationally. In order to produce a 
non-negligible PBH abundance, the curvature power spectrum must be amplified to 
values of order $\mathcal{O}(10^{-2})$ on scales far smaller than those directly 
constrained by CMB ans LSS probes. Such enhancements may originate from specific features 
in the inflationary potential, including steps~\cite{Ivanov:1994pa, 
Inomata:2017vxo}, bumps or dips~\cite{Mishra:2019pzq, Ozsoy:2018flq, 
Heydari:2023xts}, from waterfall fields in hybrid inflationary 
models~\cite{Garcia-Bellido:1996mdl, Clesse:2015wea} or even from primordial magnetic fields~\cite{Liang:2025pll, Maiti:2025ijr}. Another well-studied 
mechanism involves a transient phase of ultra-slow-roll (USR) inflation, during 
which the inflaton experiences additional friction and its velocity decreases 
rapidly~\cite{Ivanov:1994pa,Tsamis:2003px, Kinney:2005vj, Martin:2012pe}.

In conventional setups, the extra friction required to trigger the USR phase is 
typically generated either by an inflection point or a plateau in the 
inflationary potential~\cite{Garcia-Bellido:2017mdw, Germani:2017bcs, 
Motohashi:2017kbs}, by non-canonical kinetic terms~\cite{Kamenshchik:2018sig, 
Papanikolaou:2022did, Heydari:2023rmq} or through explicit modifications of the 
friction term~\cite{Romano:2016gop, Dimopoulos:2017ged}. In this work, we focus 
on the last possibility, considering USR inflation induced by modified friction 
of gravitational origin, while remaining within a single-field inflationary framework and 
avoiding the introduction of sharp features or discontinuities in an otherwise 
smooth inflationary potential.

Our motivation to work within modified gravity theories arises from both 
theoretical and phenomenological considerations \cite{CANTATA:2021asi}. From a 
theoretical perspective, modified gravity frameworks offer ways to address 
issues such as spacetime singularities and renormalizability issues that arise in 
general relativity \cite{Clifton:2011jh, Nojiri:2010wj}. From an observational 
standpoint, modifications of gravity can effectively describe dark components of 
the Universe, including dark energy and dark matter, either through genuine 
gravitational degrees of freedom or through effective energy-momentum 
contributions~\cite{Capozziello:2011et,Cai:2015emx}, and moreover can 
alleviate various cosmological tensions \cite{CosmoVerseNetwork:2025alb}. Among 
such classes, the Horndeski (or Generalized Galileon) gravity occupies a 
distinguished position, as it represents the most general scalar-tensor theory 
in four dimensions with a single scalar field, leading to second-order field 
equations and avoiding Ostrogradsky instabilities~\cite{Horndeski:1974wa, 
Deffayet:2011gz, Kobayashi:2011nu}. The explicit dependence of the Lagrangian on 
the canonical kinetic term of the scalar field makes this framework particularly 
relevant in describing inflationary and dark energy epochs \cite{ 
Deffayet:2013lga,Kase:2014yya,Tsujikawa:2014uza,Arroja:2015yvd, 
 Bellini:2015xja,DeFelice:2015isa,Babichev:2016rlq,BenAchour:2016cay, 
Pogosian:2016pwr,Arroja:2017msd,Ijjas:2017pei, 
 Capozziello:2018gms,Frusciante:2018jzw,Franchini:2019npi, 
 Heisenberg:2019qxz,Destounis:2019omd,Ilyas:2020qja, 
Oikonomou:2020sij,Dialektopoulos:2021ryi, 
Petronikolou:2021shp,Takahashi:2022mew, Horndeski:2024sjk}.

Primordial black hole formation in Horndeski gravity has been explored in 
several recent works, in order to induce a USR phase~\cite{Fu:2019ttf, 
Dalianis:2018frf, Lin:2020goi}. However, the observation of the binary neutron 
star merger GW170817 imposes stringent constraints on scalar-tensor theories, 
requiring the speed of gravitational waves to coincide with the speed of light 
at late times. This condition severely restricts the allowed Horndeski 
interactions, effectively excluding higher-order Galileon 
terms~\cite{TheLIGOScientific:2017qsa, Creminelli:2017sry, Ezquiaga:2017ekz}. 
Within these GW170817-compatible Horndeski models, enhancements of the 
curvature power spectrum have so far been achieved mainly by engineering the dependence of Hordneski terms on the inflaton field $\phi$~\cite{Yi:2020cut, Zhang:2021vak}. 

In this paper, we depart from this standard approach. While maintaining a 
canonical kinetic term and a smooth inflationary potential, we allow for a 
nontrivial dependence of the cubic Horndeski function on the canonical kinetic 
term. This kinetic structure modifies the effective friction acting on the 
inflaton and the dynamics of curvature perturbations, enabling a transient 
ultra-slow-roll phase without the need for sharp potential features. As a 
result, the growth and localization of the peak in the curvature power spectrum 
can be efficiently controlled, leading to PBH production in mass ranges 
consistent with current observational constraints. In particular, we 
demonstrate that PBHs with asteroid-scale masses can form and constitute a 
substantial fraction of the present dark matter abundance within this framework.

The paper is organised as follows. In \Sec{sec:Horndeski}, we introduce the 
Horndeski gravity setup considered in this work. In \Sec{sec:USR_Inflation}, we 
analyse the realization of ultra-slow-roll inflation, discussing separately the 
background dynamics and scalar perturbations. In \Sec{sec:PBH}, we study the 
resulting curvature power spectrum and the corresponding primordial black hole 
abundance. Finally, \Sec{sec:Conclusions} is devoted to our conclusions.

 \section{The fundamentals of Horndeski gravity and 
cosmology}\label{sec:Horndeski}

In this section we briefly review  Horndeski gravity. The action of the theory has the following form: \cite{Horndeski:1974wa, 
Deffayet:2011gz, Kobayashi:2011nu}
\begin{equation}
S=\int d^{4}x\sqrt{-g}\left({\cal L}+{\cal L}_m\right),\,\label{action1}
\end{equation}
where
\begin{equation}
{\cal L}=\sum_{i=2}^{5}{\cal L}_{i},\label{Lagsum}
\end{equation}
 with
\begin{align}
&{\cal L}_{2} = K(\phi,X),\label{eachlag2}\\
&{\cal L}_{3} = -G_{3}(\phi,X)\Box\phi,\\
&{\cal L}_{4} = G_{4}(\phi,X)\,
R+G_{4,X}\,[(\Box\phi)^{2}-(\nabla_{\mu}\nabla_{\nu}\phi)\,(\nabla^{\mu}
\nabla^{\nu}\phi)]\,,\\
&{\cal L}_{5} = G_{5}(\phi,X)\,
G_{\mu\nu}\,(\nabla^{\mu}\nabla^{\nu}\phi)\,\nonumber\\&\ \ \
\ \ \ \ -\frac{1}{6}\,
G_{5,X}\,[(\Box\phi)^{3}-3(\Box\phi)\,(\nabla_{\mu}\nabla_{\nu}\phi)\,
(\nabla^{\mu}\nabla^{\nu}\phi)
\,\nonumber\\&\ \ \
\ \ \ \ \ \ \ \ \ \ \ \ \ \ \ \ \ 
+2(\nabla^{\mu}\nabla_{\alpha}\phi)\,(\nabla^
{\alpha}\nabla_{\beta}\phi)\,(\nabla^{\beta}\nabla_{\mu}\phi)],\,\label{eachlag5
}
\end{align}
where ${\cal L}_m$ stands for the Lagrangian describing the matter content of 
the universe. In the expressions for $\mathcal{L}_i$,  $X \equiv -\frac{1}{2} 
g^{\mu\nu} \partial_\mu \phi \partial_\nu \phi$ denotes the canonical kinetic 
term, $R$ the Ricci scalar and $G_{\mu\nu}$ the Einstein tensor. 
Furthermore, the function $K(\phi,X)$ embodies a generalization of the scalar 
energy contribution, while $G_3(\phi,X)$ can be viewed as a generalized 
derivative $\phi$-dependent coupling between the metric tensor and the 
second-order derivatives of $\phi$, inducing a scalar-metric kinetic mixing 
known as kinetic gravity braiding. Additionally, $G_{4}$ is a generalized 
$G_{5,X}$ terms respectively introduce  quartic and  quintic interactions of 
$\phi$ due to the kinetic dependence of $G_4$ and $G_5$. 

In order to apply Horndeski gravity at a cosmological framework, we introduce 
the flat Friedmann-Lemaître-Robertson-Walker (FLRW) metric
\begin{equation}
\mathrm{d}s^2 = 
-\mathrm{d}t^2 + a^2 (t) (\mathrm{d}x^2 + \mathrm{d}y^2 + \mathrm{d}z^2),
\end{equation}
where $a(t)$ is the scale factor. Hence, in this case the scalar field will be 
just a function of the cosmic time, namely $\phi(t)$, while its kinetic energy 
 becomes $X=\dot{\phi}^2/2$.
 
As we mentioned in the Introduction, in the present work we are interested in  
Horndeski subclasses that are consistent with GW170817 observations. Since the  
gravitational-wave 
speed in Horndeski theories around an FLRW background is in general   different 
than 1, and it is given by  \cite{DeFelice:2011bh}
\begin{equation}
c_{T}^{2}\equiv\frac{ 2G_{4}-2XG_{5,\phi}\!-\!2XG_{5,X}\ddot{\phi} }{ 2 
(G_{{4}}-2 
XG_{{4,X}})-2X (G_{{5,X}}{\dot{\phi}}H-G_{{5,\phi}})}
\geq0,
\label{cTcon2}
\end{equation}
we can see that excluding $G_{4X}$ and 
$ G_{5}$ contributions immediately   ensures viability against GW170817 
contraints. Thus,  the action of our theory remains as 
\begin{equation}\label{eq:S}
S = \int \mathrm{d}^4x \sqrt{-g} \left[ K(\phi, X) - G_3(\phi, X)\Box\phi  
+G_4(\phi)R  \right].
\end{equation} 
We choose $G_4=\Mp^{2}/2$ to directly recover the Einstein-Hilbert term, and 
consider the standard form for the function $K$ which is $K = X - V(\phi)$, 
where $V(\phi)$ is the scalar potential. We observe that  this choice retains 
the canonical kinetic term and standard slow-roll behavior in the limit $G_3 
\to 0$, allowing at the same time for effects of kinetic self-interactions to 
be systematically studied through the function $G_3(\phi, X)$.

Working at the background level, one can extract the respective Friedmann and 
Klein-Gordon equations for the dynamical evolution of the scale factor and the 
scalar field $\phi$ by minimising respectively the action \eqref{eq:S} with 
respect to the metric $g_{\mu\nu}$ and the scalar field $\phi$. 
In particular, the  Friedmann equations in Planck units read as
\cite{Horndeski:1974wa, 
Deffayet:2011gz, Kobayashi:2011nu} 
\begin{align}\label{eq:Friedmann}
     3H^2 &=V+ \frac{\dot{\phi}^2}{2}+\dot{\phi}^2 (3H\dot{\phi}G_{3X} 
-G_{3\phi}), \\ \label{eq:Friedmann2}
     2\dot{H}+3H^2 &= V-\frac{\dot{\phi}^2}{2}\left[1 -2(\ddot{\phi}G_{3X} 
+G_{3\phi})\right],
\end{align}
where  $H\equiv 
\dot{a}/a$ is the   Hubble function.
Additionally,  the
$\phi$-evolution equation is extracted as
\begin{eqnarray}\label{eq:KG_phi}
    \ \ddot{\phi}+  3H\dot{\phi} \left(\frac{{\cal{A}}}{{\cal{C}}}\right) 
+\frac{{\cal{B}}}{{\cal{C}}}         =0 ,
\end{eqnarray}
with 
\begin{eqnarray}
&&
\!\!\!\!\!\!\!\!\!\!\!\!\!\!\!\!\! 
{\cal{A}}=1+ H\dot{\phi}G_{3X}[3-\epsilon] 
-2G_{3\phi}+\dot{\phi}^2G_{3\phi 
X} \nonumber\\
&&\!\!\!\!\!\!\!\!\!\!\!\!\!\!\!\!\!\! 
{\cal{B}}= V_{\phi}-\dot{\phi}^2G_{3\phi\phi}
\nonumber\\
&&\!\!\!\!\!\!\!\!\!\!\! \!\!\!\!\!\!  {\cal{C}}=1+6H\dot{\phi}G_{3X} 
-2G_{3\phi}+3H\dot{\phi}^3G_{3XX}-\dot{\phi}^2G_{3\phi X}.
\end{eqnarray}

We proceed by examining the linear scalar perturbations. Working in the unitary 
gauge, namely setting  
$\delta\phi=0$,   we perturb the spatial part of the metric as 
$\gamma_{ij}=a^2(t)  e^{2\zeta} (e^{h})_{ij}$, where $\zeta$ is the curvature 
perturbation and $h_{ij}$  the tensor 
perturbation 
\cite{DeFelice:2011bh,Kobayashi:2011nu,Akama:2018cqv}. The importance of the curvature perturbation $\zeta$ lies
in the fact that, on super-horizon scales, $\zeta$ is conserved~\cite{Wands:2000dp} and, thus, it can be used to propagate intact the inflationary power spectrum from the end of inflation onwards. To extract then the equation of motion of the scalar curvature perturbation we need to derive the quadratic action 
in $\zeta$, ultimately recast as \cite{Kobayashi:2011nu}
\begin{equation}\label{eq:S_pert_1}
    S^{(2)}=\int \mathrm{d}t \mathrm{d}^3x a^3 \bigg(\mathcal{G}_s 
\dot{\zeta}^2  -\frac{\mathcal{F}_s}{a^2} (\nabla \zeta)^2 \bigg),
\end{equation}
where  $\mathcal{G}_s$, $\mathcal{F}_s$ are   given in the Appendix.  
 Working in terms of the conformal time $\eta$ defined as $d\eta=\mathrm{d}t/a$ 
and introducing a new curvature perturbation variable, usually called 
Mukhanov-Sasaki (MS) variable $u$, defined as $u\equiv z_s \zeta$ where 
$z_s=\sqrt{2\mathcal{G}_s}a$, one can write  (\ref{eq:S_pert_1}) as

\begin{equation}\label{eq:S_pert_2}
    S^{(2)}=\frac{1}{2}\int \mathrm{d}\eta \mathrm{d}^3x \bigg(u'^2 -c_s^2 
(\nabla u)^2 +\frac{z_s''}{z_s} u^2 \bigg),
\end{equation}
where $c_s=\mathcal{F}_s/\mathcal{G}_s$ is the sound speed.
Hence, by minimising (\ref{eq:S_pert_2})  one extracts the equation of 
motion for the MS variable, reading as 
\begin{equation}\label{eq:MS_equation_eta}
   u_{k}'' +\bigg(c_\mathrm{s}^2 k^2 -\frac{z_s''}{z_s} \bigg)u_k =0.
\end{equation}
Moreover,  in terms of the e-fold number defined as $N\equiv \ln a$,   
(\ref{eq:MS_equation_eta}) can be recast as
\begin{equation}
    u_{k}'' + \bigg(1 +\frac{H'}{H} \bigg) u_{k}' + \Bigg\{ \frac{c_\mathrm{s}^2 
k^2}{a^2 H^2} -\frac{1}{z_s} \bigg[z_s'' + \bigg(1 +\frac{H'}{H}\bigg)z_s' 
\bigg] \Bigg\} u_{k} =0,
\end{equation}
where $H$ is the usual (cosmic-time) Hubble function, primes now denote 
derivatives with respect to $N$, and the quantities $z_s'/z_s$ and  
$z_s''/z_s$ read as 
\begin{align}
    \frac{z_s'}{z_s} &=1 +\frac{1}{2}\frac{\mathcal{G}_s'}{\mathcal{G}_s} 
\nonumber \\
    \frac{z_s''}{z_s}&= \frac{1}{2}\frac{\mathcal{G}_s''}{\mathcal{G}_s} + 
\frac{\mathcal{G}_s'}{\mathcal{G}_s}+1 
-\frac{1}{4}\bigg(\frac{\mathcal{G}_s'}{\mathcal{G}_s} \bigg)^2.
\end{align}

\section{Ultra-slow roll inflation in Horndeski gravity}
\label{sec:USR_Inflation}

In this section we investigate how ultra-slow-roll inflation can be realized 
within the GR - and GW170817-compatible Horndeski framework introduced above. We analyze 
separately the background dynamics and the evolution of scalar perturbations, 
emphasizing the role of X-dependence of the cubic Horndeski interaction in inducing a transient 
enhancement of friction without introducing  peculiar features neither in the $\phi$- nor $X$-dependence of the model.

\subsection{Background dynamics and enhanced friction}
\label{subsec:USR_background}

Focusing now on an inflationary realization within our Horndeski gravity 
framework, we introduce the first Hubble-flow slow-roll parameter $\epsilon$, 
defined as
\begin{equation}
\epsilon \equiv -\frac{\dot{H}}{H^2}.
\end{equation}
Making use of the Friedmann equations~\eqref{eq:Friedmann} and~\eqref{eq:Friedmann2}, one can show that 
$\epsilon$ takes the form
\begin{equation}
\epsilon = \frac{\dot{\phi}^2}{2H^2}
\left[ 1 - G_{3X}(\ddot{\phi} - 3H\dot{\phi}) - 2G_{3\phi} \right].
\end{equation}

In what follows, we decompose for simplicity the cubic Galileon function as
\begin{equation}
G_3(\phi,X) = f(\phi)\, g(X).
\end{equation}
In order to realize a standard slowly varying phase, namely 
with $\dot{\phi}=\sqrt{2X}$ remaining small, we exclude $X$-dependences of the 
type $1/\dot{\phi}^n$ with $n \geq 1$, since these $X$-dependences would render the 
$\dot{\phi}$-dependent terms in~\eqref{eq:Friedmann}-\eqref{eq:KG_phi} 
excessively large, leading to a breakdown of the slow-roll regime. 

Furthermore, since the USR phase is expected to occur when 
$X$ becomes very small, we also avoid monotonic dependences such as 
$\dot{\phi}$, $\ln\dot{\phi}$, or $1/\dot{\phi}^n$ with $n<1$, which would 
suppress the contribution of $G_3$ and prevent any enhanced friction from 
developing. Exploiting instead the smallness of $X$ during USR, we adopt an 
exponential dependence
\begin{equation}
g(X) = A e^{-nX},
\end{equation}
where $A$ and $n>0$ are constants. This choice allows the system to remain in a 
standard slow-roll phase while enabling a transient enhancement of the $G_3$ 
contribution when $X$ becomes sufficiently small.

Under the assumptions $|\epsilon|,|\eta|,|\beta| \ll 1$, with
\begin{equation}
\eta = -\frac{\ddot{\phi}}{H\dot{\phi}}, \qquad
\beta = \frac{\dot{\phi}^2 G_{3\phi\phi}}{V_\phi},
\end{equation}
slow-roll inflation is governed by the approximate Friedmann equations
\begin{eqnarray}
\label{Fr1approx}
&&3H^2  \simeq V, \\
&&3H\dot{\phi}(1+D_1+D_2) + V_\phi \simeq 0, 
\label{Fr2approx}\\
&&\epsilon \simeq \frac{\dot{\phi}^2}{2H^2}(1+D_1+D_2),
\label{epsapprox}
\end{eqnarray}
where
\begin{eqnarray}
&&D_1(\phi,X) = -3nAH\dot{\phi}f e^{-nX}
= 3H\dot{\phi}G_{3X}, \nonumber\\
&& D_2(\phi,X) = -2Af_\phi e^{-nX}
= -2G_{3\phi}.
\end{eqnarray}
In practice, we will assume $D_1 \ll D_2$, which simplifies the analysis 
without affecting the qualitative behavior. Having extracted above the Friedmann, the Klein-Gordon equations and the first slow-roll parameter in the slow-roll regime, one can compute the power spectrum of the curvature perturbation $\zeta$ \cite{Kobayashi:2011nu}. After a straightforward calculation, one gets that during the slow-roll phase
\begin{equation}
    \mathcal{P}_\zeta \simeq \frac{H^4}{4\pi^2 \dot{\phi}^2(1 + D_2)} \simeq \frac{V^3}{12\pi^2 V_\phi^2 }(1 + D_2), \label{eq:20}
\end{equation}
while the power spectrum of first-order tensor perturbations remains unchanged, that is, $P_T = 2H^2/\pi^2 \simeq 2V /3\pi^2$. Following this, the scalar spectral index and the tensor-to-scalar ratio will be given by the following approximated expressions:  
\begin{align}
    n_S - 1 &\simeq \frac{2}{1 + D_2} \left( \eta_V - 3\epsilon_V + \frac{D_{2\phi}}{1 + D_2} \sqrt{\frac{\epsilon_V}{2}} \right), \label{eq:21}\nonumber \\
    r &\simeq \frac{16 \epsilon_V}{1 + D_2},
\end{align}
where $\epsilon_V = (V'/V)^2/2$ and $\eta_V = V''/V$ are the potential slow-roll parameters.

While the kinetic dependence of $G_3$ determines the enhancement of friction, 
the onset and exit of the USR phase, as well as its localization toward the end 
of inflation, are controlled by the $\phi$-dependence of $f(\phi)$. To this 
end, motivated by its relation with the power spectrum in
expression~\eqref{eq:20}, we choose

\begin{equation}
f(\phi) = -
B C \sinh^{-1}\left[ \frac{\phi-\phi_c}{C} \right],
\end{equation}
which leads to a Lorentzian-type localized profile of its derivative $f_{\phi}$, where  $B$ and $C$ are constants and $\phi_c$ denotes the field value at which 
the USR phase is triggered. The parameter $B$ controls the amplitude of the 
friction enhancement, while $C$ regulates the duration of the USR phase in 
terms of e-folds. This $f$ form  ensures a smooth entrance into, and exit from, the 
USR regime.

The mechanism described above can operate for a broad class of inflationary 
potentials. For concreteness, we adopt a logarithmic potential
\begin{equation}
V(\phi) = V_0 \ln(\alpha + \gamma \phi^\delta),
\end{equation}
with $\alpha, \gamma>0$ and $\delta>0$, ensuring that the potential is smooth 
and positive for $\phi>0$. This form interpolates between monomial and 
plateau-like behavior and may arise in effective descriptions of high-energy or 
string-inspired scenarios~\cite{Baumann:2014nda,Ballesteros:2015noa,Dalianis:2018frf}. In the following, we fix $\alpha=\gamma=\delta=1$ in 
Planck units for simplicity, and we restrict to $\phi>0$, corresponding to the 
dynamically relevant inflationary region. 

Fixing $V_0$ to match CMB observations at 
$k_{\text{CMB}} = 0.05~\mathrm{Mpc}^{-1}$ and $\phi_{\rm in}=6.2$, we obtain 
$\mathcal{P}_\zeta \simeq 2.1\times 10^{-9}$, $n_s=0.9673$, and $r=0.038$. 
Achieving the enhancement necessary for PBH formation on small scales requires 
$A\cdot B = \mathcal{O}(10^8)$, with $C$ chosen sufficiently small to control 
the USR duration. In what follows, we adopt $A=10^3$ and $n=1$ as 
representative values. We note here the model's negative running of the spectral index ($\alpha_s=
\mathrm{d}n_s/\mathrm{d}\ln k<0$) which tends to be prefered by single-field USR models, as discussed in ~\cite{Allegrini:2025jha}, whereas combination of ACT and Planck data ~\cite{Aiola:2020azj} suggest a preference for positive values. This tension, which at the moment is at level of~ $2\sigma$~\cite{Allegrini:2025jha}, will be further investigated in a future study.

\subsection{Scalar perturbations and power spectrum enhancement}
\label{subsec:USR_perturbations}

During the ultra-slow-roll phase, slow-roll conditions are violated by 
construction, and the approximate expressions derived above cease to be valid. 
In this regime, the dynamics of scalar perturbations must be obtained by 
solving the MS equation, which we solve numerically, 
using~(\ref{Fr1approx})-(\ref{epsapprox})
with $D_1 \ll D_2$, as initial conditions.

For the MS variable $u_k$ and its conformal-time derivative $u_k'$, we impose 
Bunch-Davies initial conditions deep inside the horizon ($-k\eta \to \infty$),
\begin{equation}
u_k = \frac{1}{\sqrt{2c_\mathrm{s} k}}, \qquad
u_k' = -i \sqrt{\frac{c_\mathrm{s} k}{2}},
\end{equation}
where $c_\mathrm{s}$ denotes the effective sound speed.

\begin{figure}[ht!]
  \centering
  \includegraphics[width=0.9 \columnwidth]{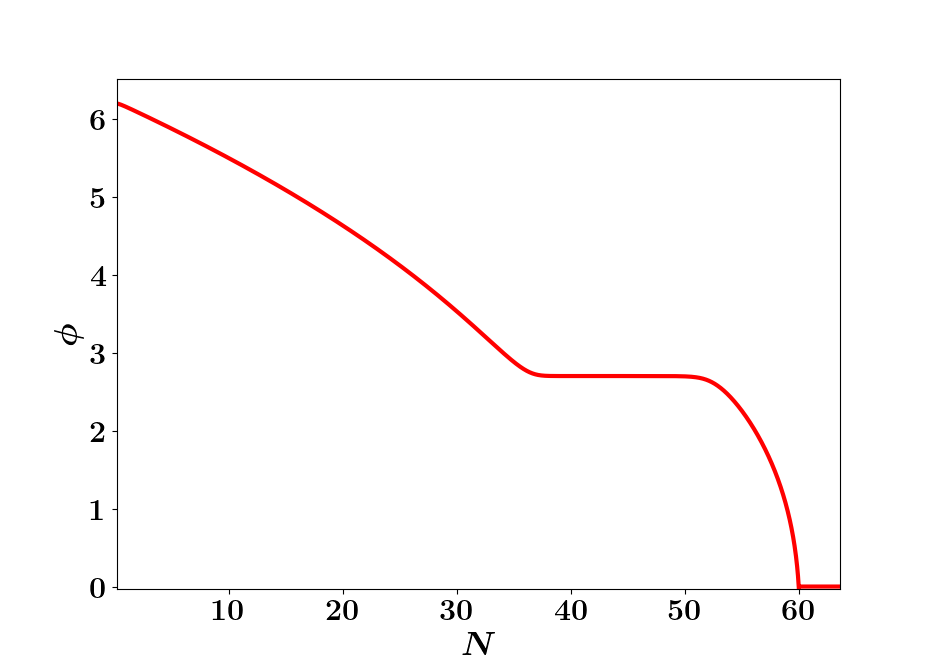}\\
  \includegraphics[width=0.9 \columnwidth]{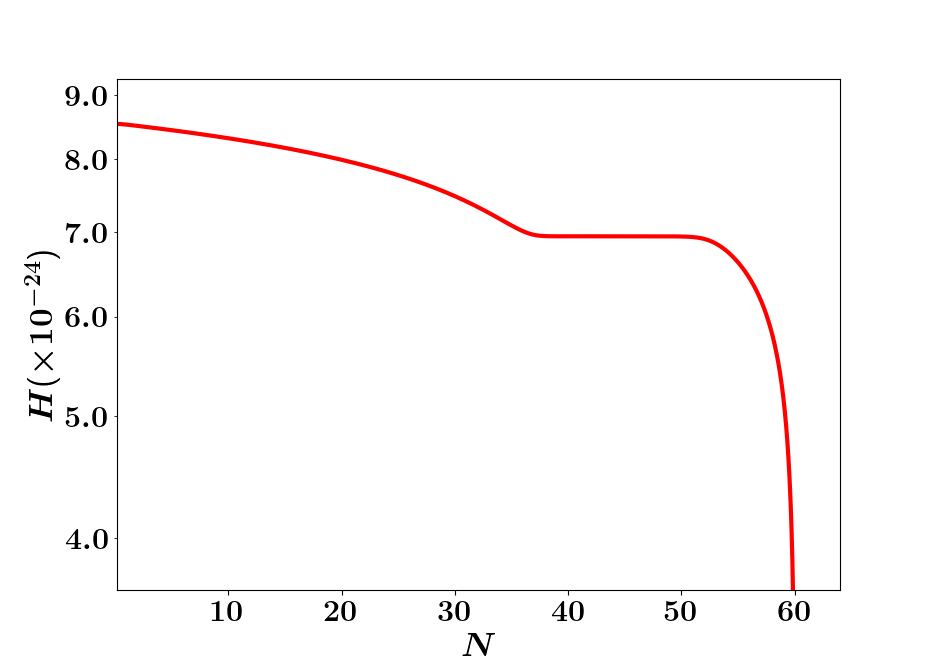}
  \caption{{\it{Evolution of the field $\phi$ and the Hubble parameter $H$ with 
  respect to the e-fold number $N$, for the case (a) of Table} \ref{tab:PBH_cases}.}}
  \label{f1}
 \end{figure}

\begin{figure}[ht!]
  \centering
  \includegraphics[width=0.9 \columnwidth]{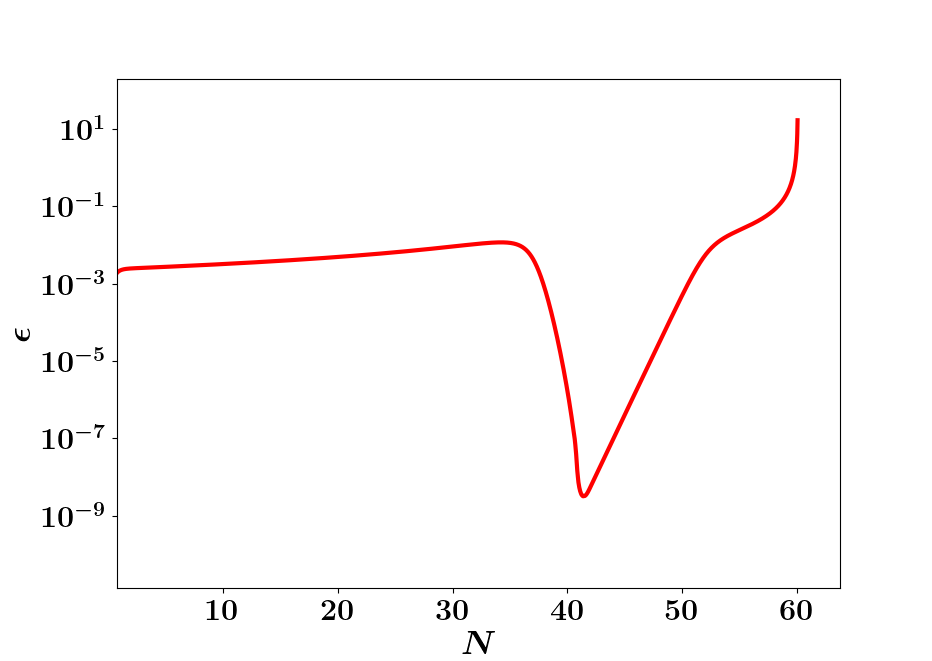}\\
  \includegraphics[width=0.9 \columnwidth]{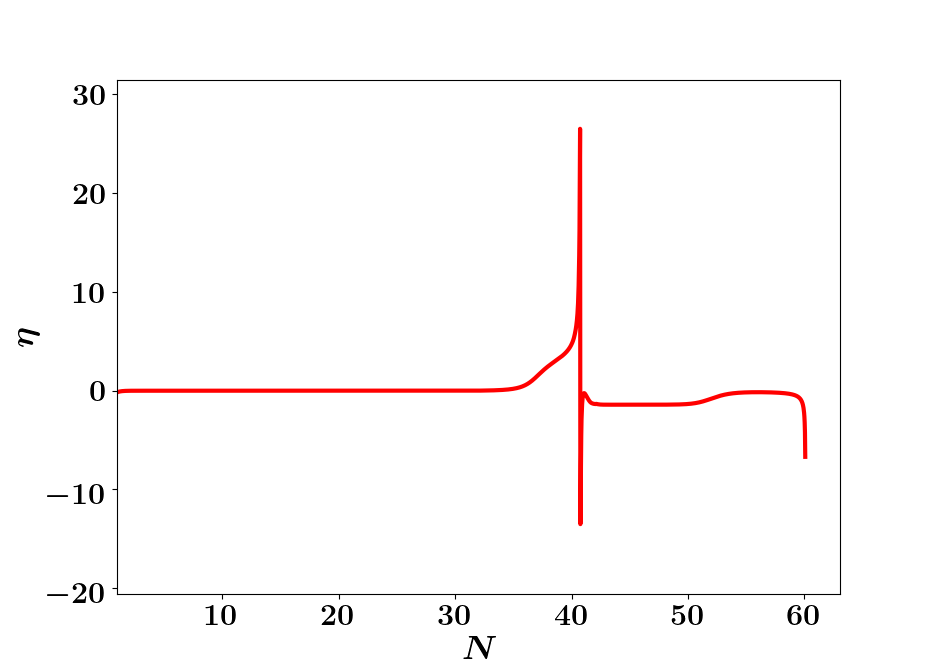}
  \caption{{\it{Evolution of the first ($\epsilon$) and second ($\eta$) 
  slow-roll parameters with respect to the e-fold number $N$, for the case (a)  
  of Table  \ref{tab:PBH_cases}.} }}
  \label{f2}
 \end{figure}

\begin{figure}[ht!]
  \centering
  \includegraphics[width=0.9 \columnwidth]{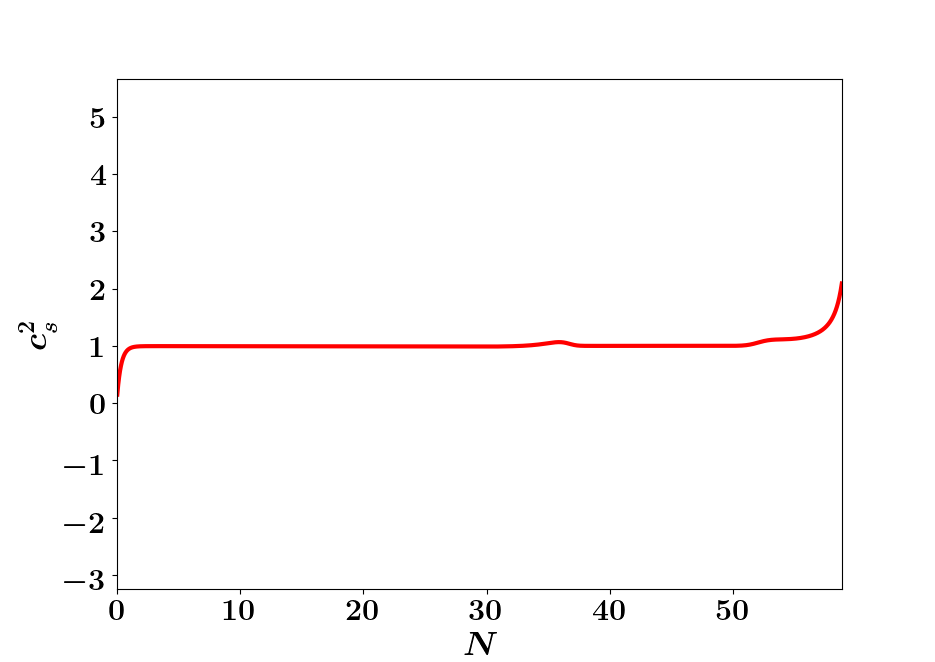}
  \caption{{\it{The sound speed squared $c_\mathrm{s}^2$ as a function of the e-fold number $N$, during the SR-USR-SR range, for the case (a)  
  of Table \ref{tab:PBH_cases}.} }}
  \label{cs2}
 \end{figure}

\begin{figure}[h!]
    \centering
    \includegraphics[width=1.1\columnwidth]{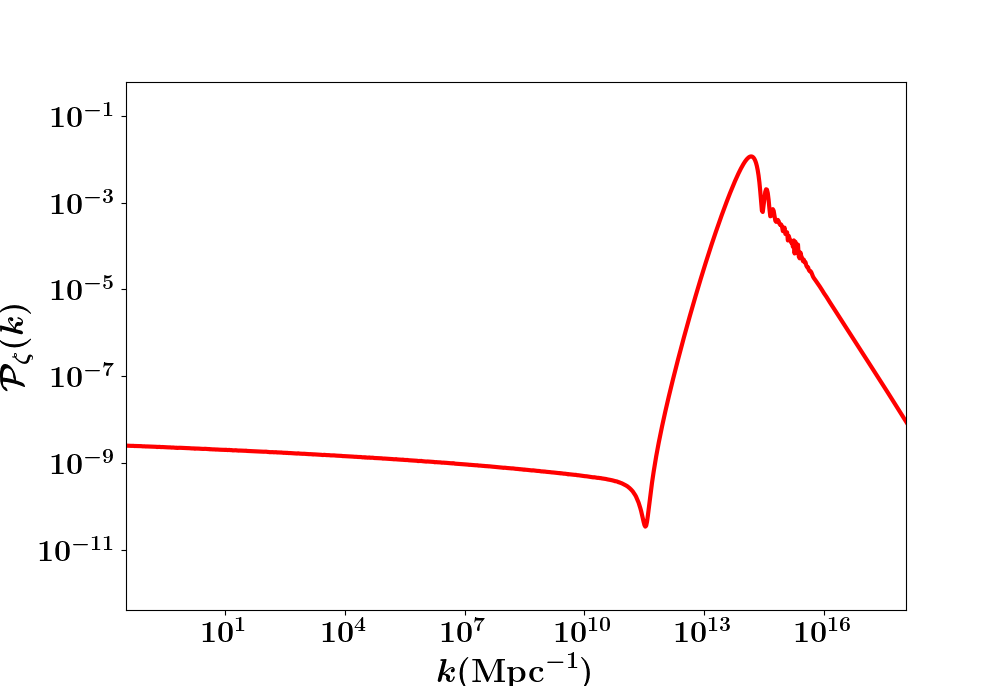}\\
  \includegraphics[width=1.1 \columnwidth]{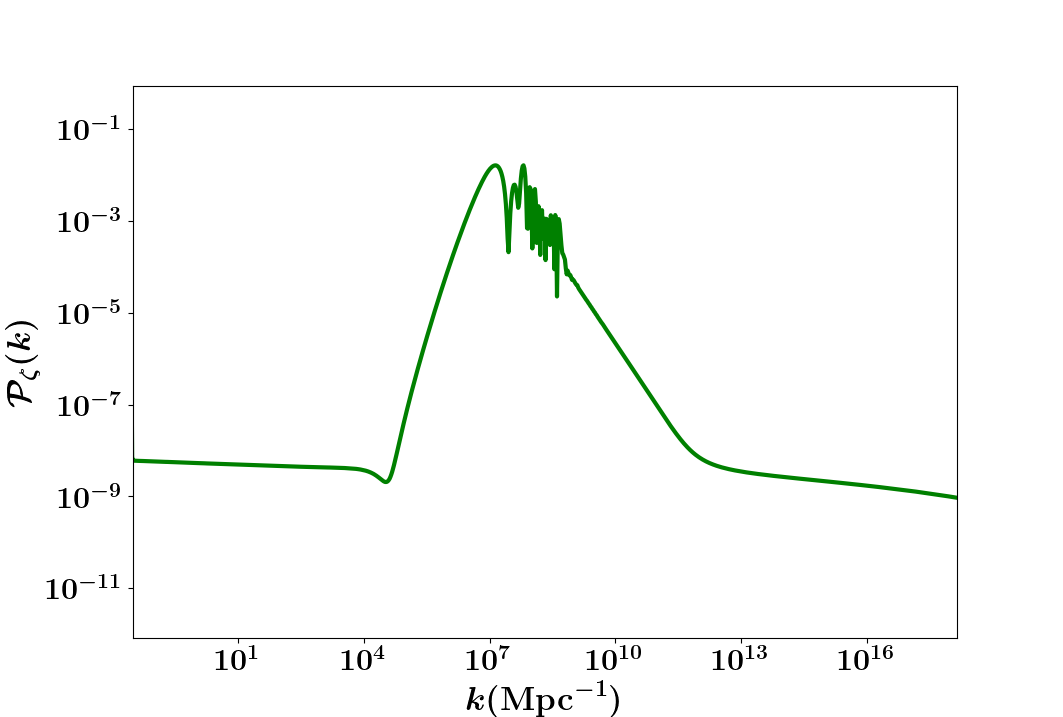}
    \caption{{\it{Profile of the scalar power spectrum $\mathcal{P}_{\zeta}$ as a function of the comoving wavenumber $k$, for the case (a) (red color) and (b) (green color)  of Table  \ref{tab:PBH_cases}.}}}
    \label{f3}
\end{figure}

We present two representative parameter choices, namely   cases (a) and (b), in 
Table~\ref{tab:PBH_cases}, which lead to qualitatively different PBH spectra. For both cases, quantities $\mathcal{F}_s$ and $\mathcal{G}_s$ remain positive, so there are no ghost or gradient instabilities, and additionally other viable configurations also
exist. In Figs.  \ref{f1}, 
\ref{f2}, \ref{f3} we show the background 
evolution and the scalar power spectrum for case (a). The USR phase occurs around 
$N\simeq 35$-$50$ e-folds, while standard slow-roll inflation is recovered 
outside this interval and ends at $N\simeq 59$. In Fig. \ref{cs2}, we also show the squared sound speed behavior $c_\mathrm{s}^2$ for the case (a). As one may infer, $c_\mathrm{s}^2$ remains positive, continuous and smooth throughout the relevant inflationary evolution, denoting that no gradient instabilities develop and perturbation equations remain well-defined.

\begin{table*} 
\centering
\caption{Representative parameter values yielding an enhancement of the scalar
power spectrum consistent with observational constraints and leading to
PBH production.
All dimensional quantities are expressed in Planck units.}
\label{tab:PBH_cases}
\renewcommand{\arraystretch}{1.15}
\begin{tabular}{|c|c|c|c|c|c|c|c|c|c|}
\toprule\hline
Case & $V_0$ $(\times10^{-9})$ &
$\phi_{\mathrm{c}}$ &
$B$ $(\times10^{5})$ &
$C$ $(\times10^{-10})$ &
$n_{\mathrm{s}}$ &
$k_{\mathrm{peak}} (\mathrm{Mpc}^{-1})$ &
$P_{\zeta,\mathrm{peak}}$ $(\times10^{-2})$ &
$M_{\mathrm{PBH}}/M_{\odot}$ &
$f_{\mathrm{PBH}}$ \\
\midrule\hline
(a) & $0.656 $ &
2.70 &
$3.85 $ &
$1.52 $ &
0.9673 &
$1.51\times10^{14}$ &
$1.16 $ &
$1.98\times10^{-16}$ &
0.95 \\

(b) & $1.90 $ &
4.82 &
$2.09 $ &
$1.40 $ &
0.9690 &
$6.45\times10^{7}$ \& &
$1.62 $ \& &
$6.35\times10^{-4}$ \& &
$4.43\times 10^{-6}$  \& \\
 &  &
 &
 &
&
 &
$1.41\times 10^{7}$ ($\mathrm{secondary}$) &
$1.62 $ ($\mathrm{sec.}$) &
$1.26\times10^{-2}$ ($\mathrm{sec.}$) &
$2.81\times 10^{-3}$ ($\mathrm{sec.}$) \\
\bottomrule\hline
\end{tabular}
\end{table*}

As we observe from Fig. \ref{f2}, the rapid suppression of $\dot{\phi}$ during USR is reflected in 
the deepening of the slow-roll parameter $\epsilon$ and the sharp variation of 
$\eta$, signaling the importance of $\ddot{\phi}$ in this regime. The resulting 
scalar power spectrum exhibits a pronounced peak at small scales around 
$k\simeq 10^{14}\,\mathrm{Mpc}^{-1}$, while remaining nearly scale-invariant on  CMB scales. The spectrum satisfies as well current constraints from PTA observations, 
BBN, and CMB $\mu$-distortions.

The pronounced oscillations in the power spectrum arise from the sharp, non-adiabatic nature of the transition into the USR phase.
In this framework, the kinetic structure of $G_3(X)$ introduces cumulative $\dot{\phi}$-dependent effects—absent in more conventional treatments—that modulate this behavior. Physically, modes on the left of the peak ($k<k_{\mathrm{peak}}$) exited the horizon prior to this transition; having already frozen out, they remain unaffected and smooth. Conversely, modes on the right ($k>k_{\mathrm{peak}}$) are still deep inside the horizon during the transition. These sub-horizon modes experienced the sharp change in the background dynamics while oscillating, inducing phase shifts that result in the observed constructive and destructive interference pattern. In case (b), shown in Fig.  \ref{f3}, the transition occurs earlier and intercepts larger modes during their horizon exit, extending these oscillations to the left of the peak.

 \begin{figure}[h!]
    \centering
    \includegraphics[width=1.1\columnwidth]{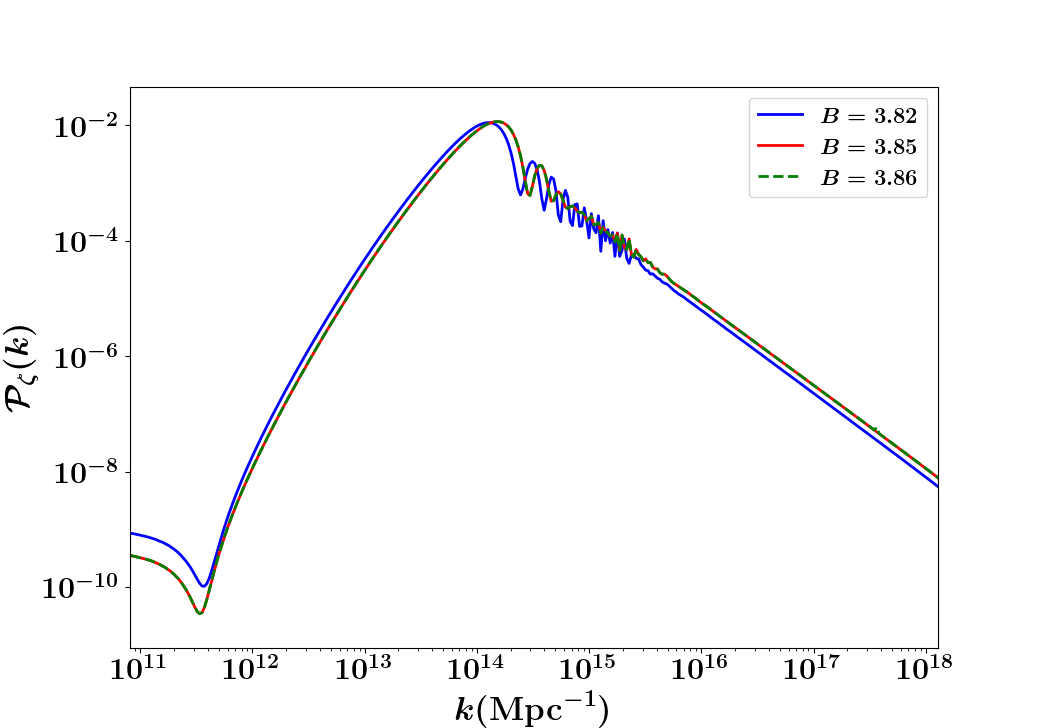}
    \caption{{\it{Profile of the scalar power spectrum $\mathcal{P}_{\zeta}$ as a function of the comoving wavenumber $k$, when changing only the $B$  parameter of the theory.}}}
    \label{tun}
\end{figure}

\section{Primordial black hole production}
\label{sec:PBH}
Let us now study the production of primordial black holes in the present 
framework, being formed from the gravitational collapse of enhanced energy density perturbations upon horizon-reentry during the radiation-dominated (RD) era after the end inflation. In order to do so we follow the peak-theory formalism~\cite{Bardeen:1985tr}.

In order to compute the abundance of PBHs from the collapse of enhanced density peaks one should relate the energy density contrast $\delta\equiv \frac{\delta\rho}{\rho_\mathrm{b}}$ with the comoving curvature perturbation $\zeta$. In the comoving gauge, we can straightforwardly show that~\cite{Mukhanov:1990me}
\begin{equation}\label{eq:zeta_vs_delta:non_linear}
\begin{split}
\frac{\delta\rho}{\rho_\mathrm{b}} &\equiv \frac{\rho(r,t)-\rho_{\mathrm{b}}(t)}{\rho_{\mathrm{b}}(t)} \\ & = -\left(\frac{1}{aH}\right)^2\frac{4(1+w)}{5+3w}e^{-5\zeta(r)(r)/2}\nabla^2e^{\zeta(r)/2},
\end{split}
\end{equation}
where $w$ is the equation-of-state (EoS) parameter $w\equiv p/\rho$. In the linear regime ($\zeta\ll 1$) the above equation will simplify to
\beq\label{eq:zeta_vs_delta:linear}
\begin{split}
\frac{\delta\rho}{\rho_\mathrm{b}}& \simeq -\frac{1}{a^2H^2}\frac{2(1+w)}{5+3w}\nabla^2\zeta(r) \\ & \Longrightarrow \delta_k =  -\frac{k^2}{a^2H^2}\frac{2(1+w)}{5+3w}\zeta_k.
\end{split}
\eeq


One should emphasize here however that the PBH gravitational collapse is a highly non-linear process. One then should account for the full non-linear relation \eqref{eq:zeta_vs_delta:non_linear} between $\delta$ and $\zeta$. Ultimately, one can show that the smoothed energy density contrast $\delta_\mathrm{m}$ is related to the linear energy 
density contrast $\delta_l$ defined through \Eq{eq:zeta_vs_delta:linear} by
~\cite{DeLuca:2019qsy,Young:2019yug}
\beq\label{eq:delta_m_smoothed}
\delta_\mathrm{m} = \delta_l - \frac{3}{8}\delta^2_l.
\eeq
Nevertheless, in order to avoid the so called cloud-in-cloud problem, namely PBH formation on small scales, the energy contrast $\delta$ should be smoothed on scales smaller than the horizon scale. At the end, the smoothed $\delta_l$ will be recast as
\beq
\delta^R_l = \int \mathrm{d}^3\vec{x}^\prime W(\vec{x},R)\delta(\vec{x}-\vec{x}^\prime),
\eeq
where the window function $W(\vec{x},R)$ is chosen to be a Gaussian window function whose Fourier transform is given by~\cite{Young:2014ana}
\beq\label{eq:Gaussian_window_function}
\tilde{W}(R,k) = e^{-k^2R^2/2},
\eeq
with $R$ (smoothing scale) being equal to the comoving horizon scale $R=(aH)^{-1}$. Making use thus of \Eq{eq:zeta_vs_delta:linear}, the smoothed variance of $\delta_l$ will read as
\beq\label{eq:sigma}
\begin{split}
\sigma^2 & \equiv \langle \left(\delta^{R}_l\right)^2\rangle = \int_0^\infty\frac{\mathrm{d}k}{k}\mathcal{P}_{\delta_l}(k,R)  \\ & = \frac{4(1+w)^2}{(5+3w)^2}\int_0^\infty\frac{\mathrm{d}k}{k}(kR)^4 \tilde{W}^2(k,R) \mathcal{P}_\zeta(k),
\end{split}
\eeq
where $\mathcal{P}_{\delta_l}(k,R)$ and $\mathcal{P}_\zeta(k)$ stand for the reduced energy density and curvature power spectra respectively.

Concerning now the PBH mass, the latter will be of the order of the cosmological horizon mass at the horizon-reentry, following ultimately a critical collapse scaling law~\cite{Niemeyer:1997mt,Niemeyer:1999ak,Musco:2008hv,Musco:2012au},  
\beq\label{eq:PBH_mass_scaling_law}
M_\mathrm{PBH} = M_\mathrm{H}\mathcal{K}(\delta-\delta_\mathrm{c})^\gamma,
\eeq
where $M_\mathrm{H}$ is the cosmological horizon mass at horizon crossing time (horizon-reentry), and $\gamma \simeq 0.36$ for PBH formation during the RD era. The parameter $\mathcal{K}$ is of the order of $\mathcal{K}\simeq 4$~\cite{Musco:2008hv} with its exact value depending on EoS parameter $w$ as well as on the shape of the collapsing overdensity.

With regard to the determination of the threshold value $\delta_\mathrm{c}$, the latter will depend of the shape of the collapsing curvature power spectrum~\cite{Musco:2018rwt}. In our case, as it can be seen from \Fig{f3} we are met with a broad curvature power spectrum. Consequently, in order to compute $\delta_\mathrm{c}$ we followed the analysis of~\cite{Musco:2020jjb} finding at the end that in our setup $\delta_\mathrm{c} = 0.3974$ independently of the choice of the parameters of the collapsing curvature spectrum. This is because $\mathcal{P}_\zeta$ has a universal shape-$k$ dependence around the peak independently of its parameters choice, namely $k^{-2.2}$ on the right region from the peak and $k^{2.3}$ on the left region from the peak.

At the end, the fraction of the Universe's energy density at a peak of a given height $\nu\equiv \delta/\sigma$, collapsing to form a PBH, $\beta_\nu$ will read 
\beq
\beta_\nu = \frac{M_\mathrm{PBH}(\nu)}{M_\mathrm{H}}\mathcal{N}(\nu)\Theta(\nu - \nu_\mathrm{c})
\eeq
while the resulting PBH mass distribution $\beta(M)$ will read as~\cite{Young:2019yug}
\beq\label{eq:beta_full_non_linear}
\beta(M) = \int_{\nu_\mathrm{c}}^{\frac{4}{3\sigma}}\mathrm{d}\nu\frac{\mathcal{K}}{4\pi^2}\left(\nu\sigma - \frac{3}{8}\nu^2\sigma^2 - \delta_{\mathrm{c}}\right)^\gamma \frac{\mu^3\nu^3}{\sigma^3}e^{-\nu^2/2},
\eeq
where $\nu_\mathrm{c} = \delta_{\mathrm{c},l}/\sigma$ and $\delta_{\mathrm{c},l}=\frac{4}{3}\left(1 -
\sqrt{\frac{2-3\delta_\mathrm{c}}{2}}\right)$.
The parameter $\mu$ stands for the first moment of the smoothed power spectrum defined as
\beq
\begin{split}
\mu^2 & =\int_0^\infty\frac{\mathrm{d}k}{k}\mathcal{P}_{\delta_l}(k,R)\left(\frac{k}{aH}\right)^2 \\ & = \frac{4(1+w)^2}{(5+3w)^2}\int_0^\infty\frac{\mathrm{d}k}{k}(kR)^4 \tilde{W}^2(k,R)\mathcal{P}_\zeta(k)\left(\frac{k}{aH}\right)^2.
\end{split}
\eeq

Finally, the present-day fraction of dark matter composed of PBHs with mass $M$ can be recast as
\begin{equation}\label{eq:f_PBH}
f_{\rm PBH}(M) =
\left(\frac{\beta(M)}{3.27 \times 10^{-8}}\right) \left(\frac{106.75}{g_{*,\mathrm{f}}}\right)^{1/4}\left(\frac{M}{M_\odot}\right)^{-1/2},
\end{equation} 
where $g_{*,f}$ denotes the effective number of relativistic degrees of freedom at 
PBH formation and $M_\odot$ is the solar mass. 

In order to have an estime of the PBH mass associated with the collapse of an overdensity region of a characteristic comving scale $k^{-1}$, one can approximate it with the cosmological horizon mass at that time. Applying
as well entropy conservation from PBH formation time up to today, the mass of a PBH formed in the RD era and associated with a comoving wavenumber $k$ will read as:
\begin{equation}\label{MKdef}
M(k) = M_\mathrm{H}\Omega^{1/2}_\mathrm{rad,0}\left(\frac{g_{*,0}}{g_\mathrm{*,f}}\right)^{1/6}\left(\frac{k_0}{k}\right)^2,
\end{equation}
where $\Omega_\mathrm{rad}=\rho_\mathrm{rad}/\rho_\mathrm{crit}$ is the radiation density parameter, and $0$ refers to current era values.
In the present analysis we take $g_{*,0} = 3.36$, $g_{*,f} = 106.75$ (as before the electroweak phase transition), and $\Omega_\mathrm{rad,0}\simeq 10^{-5}$.

\begin{figure}[h!]
    \centering
    \includegraphics[width=1.1\linewidth]{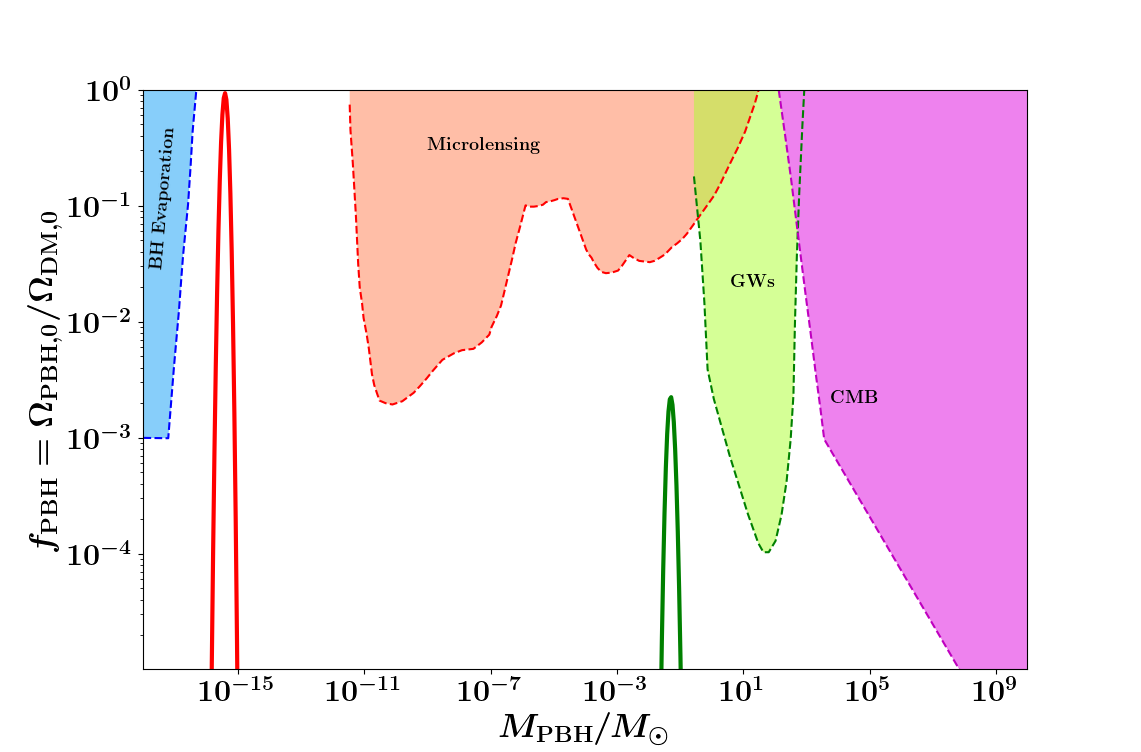} 
    \caption{{\it{The PBHs abundance $f_{\text{PBH}}$ with respect to the PBHs masses $M$, on top of the  observational contraints from CMB (pink color), GWs (greeen color), Microlensing (orange color) and BH Evaporation (blue color), for the case (a) and case (b)  of Table  \ref{tab:PBH_cases}. The asteroid-mass window, at roughly $10^{-11}-10^{-16} M_{\odot}$, could account for 100\% of DM.}}}
    \label{f4}
\end{figure}

For the curvature power spectra and the peak scales reported in 
Table~\ref{tab:PBH_cases}, we show in Fig. \ref{f4}
the PBH fraction to dark matter as a function of the PBH mass. As we can see, the resulting PBH 
mass spectrum can be broadly divided into distinct 
regimes. At sufficiently low masses, PBHs have already evaporated, or are 
evaporating today, being strongly constrained due to evaporation observational limits. On the other hand, at very large 
masses, PBHs correspond to intermediate or supermassive objects and can 
contribute only a subdominant fraction of the dark matter abundance due to 
observational constraints from microlensing~\cite{Niikura:2017zjd}, GWs~\cite{Sasaki:2016jop,Andres-Carcasona:202405732} and CMB~\cite{Ali-Haimoud:2016mbv}.

However, interestingly enough,
in between these regimes lies a mass window where PBHs can constitute a 
substantial, or even dominant, component of dark matter while remaining 
consistent with current observational bounds. In particular, for the case (a)  of Table  \ref{tab:PBH_cases} we 
obtain PBH masses of order of $M_{\rm PBH} \sim 10^{-16}M_\odot$, corresponding to 
$M \sim  10^{17}\,{\rm g}$, i.e.\ asteroid-mass PBHs. This mass range 
falls within the window where PBHs may account for essentially the entire dark 
matter abundance. In this case, the predicted PBH fraction reaches $f_{\rm 
PBH}\simeq 0.95$.
We mention here that the enhanced scalar perturbations associated with the 
ultra-slow-roll phase are also expected to source scalar-induced gravitational 
waves (SIGWs) at second order in cosmological perturbation theory~\cite{Domenech:2024drm,Domenech:2025zvi}. Given the peak scale $k_{\rm peak}\simeq10^{14}\mathrm{Mpc}^{-1}   $, the corresponding GW peak frequency $f$ can be recast as $f = k/(2\pi a_0) \sim 1\mathrm{Hz}$, where we use that today $a_0$ is normalised to unity, liying  in between the frequency detection bands of LISA and ET,
and motivating thus future dedicated studies in connection with such GW detectors.

Case (b) leads to a different PBH mass scale and abundance, resulting to a 
smaller contribution to the total dark matter density.
The intensified oscillatory behavior observed at larger mass scales modulates the power spectrum sufficiently to generate a secondary collapse window, resulting in a double-peaked PBH abundance. However, in Fig. \ref{f4}, we show only the primary peak of $f_\mathrm{PBH}$ since the seconday one is quite small, of the order of $10^{-6}$ [See also Table \ref{tab:PBH_cases}].
These examples illustrate 
how variations in the parameters controlling the ultra-slow-roll phase and the 
curvature power spectrum directly translate into distinct PBH phenomenology 
within the present framework.

At this point, let us also stress  that the location and height of the peak are sensitive to the model parameters at the level of the second decimal place, corresponding to moderate fine tuning.
To illustrate the sensitivity of PBH production on the model parameters, we vary only parameter $B$, controlling predominantly the height of the peak, while keeping all other parameters fixed. We depict thus in Fig. \ref{tun} the scalar power spectrum $\mathcal{P}_\zeta(k)$ behaviour in response to $B$ value variations. In particular, as we see from Fig. \ref{tun}, by changing $B$ at the level of the second decimal digit, $\mathcal{P}_\zeta(k)$ remains practically unchangeable. We show below how the abundance $f_\mathrm{PBH}$ changes as well in the changes of the $B$ parameter value:
\begin{equation}
(B,f_{\mathrm{PBH}})=(3.82,0.23),\quad (3.85,0.93),\quad (3.86,1.52).\nonumber
\end{equation}

Consequently, compared to conventional inflection-point or extended ultra-slow-roll realizations, where tiny parameter changes at the fifth or the nineth decimal digit  often suppress PBH production by many orders of magnitude~\cite{Germani:2017bcs,Cole:2023wyx}, the present framework achieves phenomenologically relevant abundances under comparatively moderate parameter adjustments at the level of the second decimal digit.

Finally, we mention also that ultra-slow-roll phases are known to potentially 
generate sizable non-Gaussianities in the curvature perturbations~\cite{Martin:2012pe,Atal:2018neu, Ballesteros:2024pbe}, which may 
affect primordial 
black hole abundance estimates based on Gaussian statistics. The quantitative 
impact of such effects depends sensitively on the detailed dynamics of the 
ultra-slow-roll phase as well as on the shape of the curvature power spectrum around 
its peak. A dedicated analysis of non-Gaussian corrections to the PBH abundance 
within the present Horndeski framework is beyond the scope of this work and is 
left for future investigation.
 \section{Conclusions}
\label{sec:Conclusions}

In this work  we studied the production of primordial black holes within an 
inflationary scenario embedded in the Horndeski modified gravity framework, 
imposing two key physical requirements: the propagation speed of gravitational 
waves coincides with the speed of light at late times, as dictated by the 
GW170817 observation, and the coupling to the curvature scalar remains constant, 
corresponding to the standard Einstein-Hilbert term. Under these conditions, 
and retaining a canonical kinetic term as well as a smooth inflationary 
potential 
through $K(\phi,X)$, the inflationary dynamics and the enhancement of curvature 
perturbations are governed entirely by the cubic Horndeski contribution 
$G_3(\phi,X)$.

We showed that a suitable kinetic dependence of $G_3$ can induce a transient 
phase of ultra-slow-roll inflation by effectively enhancing the friction acting 
on the scalar field, without introducing specialized features in the $\phi$-dependence of $G_{3}$ or discontinuities in the 
potential. Among standard classes of $X$-dependences, we found that an 
exponential dependence is uniquely capable of preserving a conventional 
slow-roll phase, while generating a controlled ultra-slow-roll epoch. In this 
way, even a mild, localized $\phi$-dependence in $G_3$ can be amplified into a 
pronounced peak in the curvature power spectrum, sufficient for primordial black 
hole formation, while remaining consistent with observational constraints from 
the CMB, pulsar timing arrays, big-bang nucleosynthesis, and CMB 
$\mu$-distortions. 

Applying this mechanism to a representative smooth plateau-type potential, here 
chosen in logarithmic form, we identified stable parameter configurations that 
lead to the formation of primordial black holes with masses of order 
$\mathcal{O}(10^{-16})\,M_\odot$, corresponding to asteroid-mass PBHs. This mass 
range lies just above the threshold for complete Hawking evaporation at the 
present epoch and falls within a window where primordial black holes can 
constitute a substantial fraction of the dark matter abundance while satisfying 
the other existing observational bounds. In the 
examples presented, the resulting PBH abundance can reach values close to the 
total dark matter density, $f_{\rm PBH}\simeq 0.9$. Additional viable solutions 
exist in the parameter space, leading to different PBH mass scales and 
abundances, illustrating the flexibility of the mechanism.

A notable feature of the present framework is that the required enhancement of 
the curvature power spectrum is achieved with relatively moderate parameter 
tuning compared to standard inflationary type PBH scenarios~\cite{Cole:2023wyx}, which nevertheless have been argued to satisfy Wilson’s criterion of technical naturalness~\cite{Iovino:2025tcv}.
The key 
ingredient is the kinetic structure of the Horndeski interaction rather than a 
finely engineered inflationary potential or $\phi$-  dependence, allowing the background inflationary 
predictions on CMB scales to remain essentially unchanged.

The results presented here open several directions for further investigation. A 
particularly important extension concerns the stochastic gravitational-wave 
background induced at second order by the enhanced scalar perturbations 
associated with the ultra-slow-roll phase~\cite{Domenech:2024drm}. Given the sharp peak in the curvature 
power spectrum and the oscillatory pattern on high-$k$, the corresponding scalar-induced GW signal is expected to exhibit characteristic spectral features, potentially falling 
within the sensitivity ranges of future detectors such as the LISA or the Einstein 
Telescope. A detailed analysis of this signal would provide an additional 
observational handle on the mechanism discussed here. Further studies could also 
address the role of non-Gaussianities   generated during the ultra-slow-roll 
phase, as well as the embedding of the present scenario within more complete 
models of reheating and post-inflationary evolution. Together, these extensions 
would help clarify the observational viability of primordial black holes arising 
from modified-gravity-induced inflationary dynamics, and further elucidate the 
connection between early modified gravity and the present dark matter 
content of 
the Universe.

\begin{acknowledgments} 
The authors would like to thank Charalampos Tzerefos for useful feedback and stimulating discussions as well as the 
contribution of 
the LISA Cosmology Working Group (CosWG). They aknoweldge also support from the COST 
Actions CA21136 -  Addressing observational tensions in cosmology with 
systematics and fundamental physics (CosmoVerse)  - CA23130, Bridging 
high and low energies in search of quantum gravity (BridgeQG)  and CA21106 -  
 COSMIC WISPers in the Dark Universe: Theory, astrophysics and 
experiments (CosmicWISPers). TP acknowledges the support of the INFN Sezione di Napoli through the project QGSKY as 
well as financial support by the funding program “MEDICUS” of the University of Patras. 
\end{acknowledgments}

\appendix 

\section{The coefficient functions of  
the quadratic action for  scalar perturbations
}\label{A}

In Horndeski gravity, the quadratic action for 
 scalar perturbations around an FLRW background is given 
as \cite{Kobayashi:2011nu}
\begin{equation}\label{eq:S_pert_1App}
    S^{(2)}=\int \mathrm{d}t \mathrm{d}^3x a^3 \bigg(\mathcal{G}_s 
\dot{\zeta}^2  -\frac{\mathcal{F}_s}{a^2} (\nabla \zeta)^2 \bigg),
\end{equation}
where $\zeta$ is the curvature 
perturbation. 
The coefficient functions are given by \cite{Kobayashi:2016xpl, Akama:2018cqv}
\begin{align}
{\cal F}_s &\equiv\frac{1}{a}\frac{d\xi}{dt}-{\cal F}_T,
\label{F_scon}
\\
{\cal G}_s &\equiv \frac{\Sigma }{\Theta^2}{\cal G}_T^2+3{\cal G}_T
\label{G_scon},
\end{align}
where
\begin{align}
{\cal F}_T&\equiv2 \left[ G_4-X \left(\ddot\phi 
G_{5,X}+G_{5,\phi}\right)\right],
\\
{\cal G}_T&\equiv2 \left[ G_4-2XG_{4,X}-X \left(H\dot\phi 
G_{5,X}-G_{5,\phi}\right)\right],
\end{align}
with
\begin{align}
\xi\equiv \frac{a{\cal G}_T^2}{\Theta},
\label{xi}
\end{align}
and 
 \begin{eqnarray}
\Sigma&\equiv&XK_{,X}+2X^2K_{,XX}+12H\dot\phi XG_{3,X}
\nonumber\\&&
+6H\dot\phi X^2G_{3,XX}
-2XG_{3,\phi}-2X^2G_{3,\phi X}-6H^2G_4
\nonumber\\&&
+6\Bigl[H^2\left(7XG_{4,X}+16X^2G_{4,XX}+4X^3G_{4,XXX}\right)
\nonumber\\
&&
\ \ \ \ \ \,
-H\dot\phi\left(G_{4,\phi}+5XG_{4,\phi X}+2X^2G_{4,\phi XX}\right)
\Bigr]
\nonumber\\&&
+30H^3\dot\phi XG_{5,X}+26H^3\dot\phi X^2G_{5,XX}
\nonumber\\&&
-6H^2X\bigl(6G_{5,\phi}
+9XG_{5,\phi X}+2 X^2G_{5,\phi XX}\bigr)
\nonumber\\&&
+4H^3\dot\phi X^3G_{5,XXX}\,,
\end{eqnarray}
 \begin{eqnarray}
\Theta
&\equiv&-\dot\phi XG_{3,X}+
2HG_4-8HXG_{4,X}
\nonumber\\&&
-8HX^2G_{4,XX}+\dot\phi G_{4,\phi}+2X\dot\phi G_{4,\phi X}
\nonumber\\&&
-H^2\dot\phi\left(5XG_{5,X}+2X^2G_{5,XX}\right)
\nonumber\\&&
+2HX\left(3G_{5,\phi}+2XG_{5,\phi X}\right)\,.
\label{theta}
\end{eqnarray}

\bibliography{bibliography}

\end{document}